\documentclass[showpacs,preprintnumbers,amsmath,amssymb]{revtex4}

\usepackage{graphicx}% Include figure files
\usepackage{dcolumn}% Align table columns on decimal point
\usepackage{bm}% bold math
\usepackage[pdftex,colorlinks,urlcolor=blue,filecolor=magenta]{hyperref}
\usepackage{url}
\usepackage{epsfig}

\newcommand{\be}{\begin{equation}}
\newcommand{\ba}{\begin{eqnarray}}
\newcommand{\ee}{\end{equation}}
\newcommand{\ea}{\end{eqnarray}}

\newcommand{\fr}{\frac}
\newcommand{\oo}{\omega_{0}}
\newcommand{\oa}{\omega_{a}}

\begin{document}

\title{Drawbacks of Principal component analysis}

%\\

\author{Seokcheon Lee}
\email{skylee@phys.sinica.edu.tw}
\affiliation{Institute of Physics, Academia Sinica, Nankang,
Taiwan 11529, R.O.C.}

\date{May 11, 2010}% It is always \today, today,
             %  but any date may be explicitly specified

\begin{abstract}
One of the main tasks for present and future dark energy surveys is to determine whether the dark energy is dynamical or not. To illustrate this from data, it is commonly used to parameterize the dark energy equation of state $\omega$ as several piecewise constant $\omega_{i}$s using the principal component analysis (PCA) method over finite redshift bins. We show that there is only $j-1$ free parameters $\omega_{i}$s if we choose the redshift as $j$ bins. Without this constrain, one obtains the inconsistent results from the data analysis. Furthermore, if $\omega$ decreases with non-negligible ratio as $z$ does, then PCA fails to reproduce the original behavior of $\omega$. Also, time varying $\omega$ can be confused with the incorrect value of constant one when the decreasing (or increasing) ratio of $\omega$ is small but not negligible.
\end{abstract}

% PACS, the Physics and Astronomy

\maketitle

One of the possible explanations for the current accelerating expansion of the Universe is the existence of the exotic energy budget ({\it i.e.} dark energy) in addition to the matter component. The different dark energy models are classified by the corresponding equation of state (EOS) $\omega$, defined by the ratio of its pressure and energy density. Due to our ignorance of the nature of dark energy, it might be appropriate to study the dark energy phenomenologically by using the accumulating high precision observational data. One needs to parameterize $\omega$ in order to fit the related parameters to data. One general way to parameterize $\omega$ is to approximate it using the piecewise constant bins \cite{0106079, 0207517}. It is claimed that one can reconstruct the time dependence of $\omega$ and make further model independent studies by using a principal component analysis (PCA) method. When we consider the time varying EOS, we adopt the so-called  Chevallier-Polarski-Linder (CPL) parameterization $\omega = \omega_{0} + \omega_{a} \fr{z}{1+z}$ and use its present value ($\omega_{0}$) and time variation ($\omega_{a}$) \cite{0009008,0208512}. One of the main reason for using this PCA method is known as to determine whether the dark energy density is evolving with time or not. However, we show that the time varying $\omega$ can be confused with the incorrect value of constant one when the decreasing (or increasing) ratio of $\omega$ is small but not negligible ({\it i.e.} when $\oa \le 0.3$). Thus, even if the PCA gives the similar result as the constant EOS result, we still can not rule out the time varying $\omega$ with non-negligible value of $\oa$ model. Also, when the time variation of $\omega$ is significant ({\it i.e.} when $\oa \ge 0.3$), then PCA analysis can hardly reproduce the original character of $\omega$.

We first want to emphasize the correct degree of freedom in PCA. One divides the redshift range of the survey ($z=0, z_{\rm{max}}$) into $N$ bins of not necessarily equal widths $\Delta z_i$ ($i = 1, \cdots , N$), where $\sum_i \Delta z_i = z_{\rm{max}}$. Then, it is well known that a set of $N$ values of observations of possibly correlated variables can be orthogonally transformed into a set of $j$ values of uncorrelated variables so-called principal components \cite{Pearson, 9603021}. The dark energy is parameterized in terms of $\omega(z)$, which is defined to be constant in each redshift bin, with a value $\omega_{i}$ in $i$th bin. For the piecewise constant $\omega(z)$, the energy density of the dark energy for $z$ in bin $j$ evolves as \be \rho_{\rm{DE}}(z) = \rho_{\rm{DE}}(z=0) \Biggl(\fr{1+z}{1+z_j} \Biggr)^{3(1+\omega_j)} \prod_{i=1}^{j-1} \Biggl(\fr{1+z_{i+1}}{1+z_i} \Biggr)^{3(1+\omega_i)} \, , \label{rhode} \ee where $z_i$ is the lower redshift bound of the $i$th bin and $\omega_i$ is the fiducial value of the EOS in that bin. The common mistake in literature is that one regards that there are $j$ free parameters after one fixes $z_j$s. However, we should emphasize that $\omega_i$s have only $j-1$ degree of freedom because $\rho_{DE}(z)$ should be equal to $\rho_{DE}(z=0)$ when $z=0$. From the above equation (\ref{rhode}), this is given by \be \omega_j = -1 + \Biggl( \sum_{i=1}^{j-1} (1+\omega_i) \ln \Bigl[\fr{1+z_{i+1}}{1+z_i} \Bigr] \Biggr) \Biggl/ \Bigl( \ln [ 1+z_j ] \Bigr) \, . \label{omegaj} \ee Thus, $\omega_{j}$ is determined by other parameters $\omega_i$, $z_i$, and $z_{j}$. Of course, one can constrain one of $\omega_i$s instead of $\omega_j$ without changing the final result.
%%%%%%%%%%%%%%%%%%%%%%%%%%%%%%%%%%%%%%%%%
\begin{center}
    \begin{table}
    \begin{tabular}{ | c | c | c | c | c | c | c | c | }
    \hline
      $\oo$ & $\oa$  & $z_{i}$ & $\omega_{i}$ & $\omega_{i}^{\ast}$ & $\sigma_{i}$ & $\chi^2$ & $\chi^{\ast 2}$ \\ \hline
      $$ & $$  & $0.1$ & $-0.72$ & $-0.72$ & $0.16$ & & \\ \cline{3-6}
      $$ & $$  & $0.4$ & $-0.73$ & $-0.73$ & $0.23$ & & \\ \cline{3-6}
      $-0.9$ & $0.5 $  & $0.8$ & $-0.67$ & $-0.67$ & $0.53$ & $0.60$ & $0.53$ \\ \cline{3-6}
      $$ & $$  & $1.25$ & $-0.63$ & $-0.63$ & $1.35$ & & \\ \cline{3-6}
      $$ & $$  & $1.6$ & $-0.73$ & $-0.60$ & $$ & & \\ \hline
      $$ & $$  & $0.1$ & $-1.05$ & $-1.05$ & $0.07$ & & \\ \cline{3-6}
      $$ & $$  & $0.4$ & $-1.00$ & $-1.00$ & $0.09$ & & \\ \cline{3-6}
      $-1.1$ & $0.3 $  & $0.8$ & $-0.97$ & $-0.96$ & $0.14$ & $0.062$ & $0.056$ \\ \cline{3-6}
      $$ & $$  & $1.25$ & $-0.95$ & $-0.94$ & $0.29$ & & \\ \cline{3-6}
      $$ & $$  & $1.6$ & $-1.00$ & $-0.93$ & $$ & & \\ \hline
    \end{tabular}
    \caption{Parameter values of true models are shown in the first two columns. $z_{i}$ is the uncorrelated bin and $\omega_{i}$ and $\omega_{i}^{\ast}$ are obtained from the minimum $\chi^2$ fitting by using the simulated $H(z)$ data compared to the true model with and without the constraint on the one of parameter values, respectively. $\sigma_{i}$ is the $1$-$\sigma$ error of $\omega_i$. $\chi^2$ and $\chi^{\ast 2}$ correspond to the minimum $\chi^2$ values with and without the constraint.}
    \label{table1}
    \end{table}
\end{center}
%%%%%%%%%%%%%%%%%%%%%%%%%%%%%%%%%%%%%%%%%%%%%%%%%

In what follows, we use the $40$ simulated equally binned Hubble parameter $H(z)$ data for $0 \le z < 2$ by using the true models given by CPL parametrization. We assume that the measurement error on $H$ as $5$ \% and there are no errors in both the present energy density contrast of the matter $\Omega_{m}^{0}$ and the present Hubble parameter value $H_0$. We also consider only the flat universe. We perform a simple $\chi_{H}^2$ test to determine the best fit values of $\omega_{i}$ and $1$-$\sigma$ error of each $\omega_{i}$ is obtained from the covariance matrix. We check the reliability of PCA method to constrain the DE and using $H(z)$ is good enough for this purpose. When we investigate the luminosity distance, the results are even worse because of the multi-integral of $\omega$ in it. We show the results for the two different models in table \ref{table1}. In the first case, true model is characterized by ($\oo, \oa$) = ($-0.9, 0.5$). We perform $\chi^2$ test with $H(z)$ data created from the this true model with the assumption of $5$ \% error. The true model of the second case is given by ($\oo, \oa$) = ($-1.1, 0.3$). We explain the details of the data with Fig. \ref{fig2} later. There is one remark to be emphasized. The $1$-$\sigma$ error increases as $z_{i}$ does. This is due to the fact that $\chi^2$ fitting is performed from the lowest $z_{i}$. Thus, the errors in lower $z_i$s is propagated to the higher $z_i$s. Thus, no matter how much data point we add in the higher $z_i$, this intrinsic error propagation will not be removed. We choose the uncorrelated bins as $z = 0.1$, $0.4$, $0.8$, $1.25$, and $1.6$. As we show before, there are only $4$ degree of freedom in this case. Thus, the value of $\omega_5$ at $z = 1.6$ is derived from the other $\omega_i$ and $z_{i}$ values by using Eq. (\ref{omegaj}). We compare the values of $\omega_i$ with the ones of $\omega_{i}^{\ast}$ which is obtained from $\chi^2$ test without the above constraint. As we can see there are discrepancies in the $\omega_5$ values between them.

\begin{center}
\begin{figure}
\vspace{1.5cm} \centerline{ \psfig{file=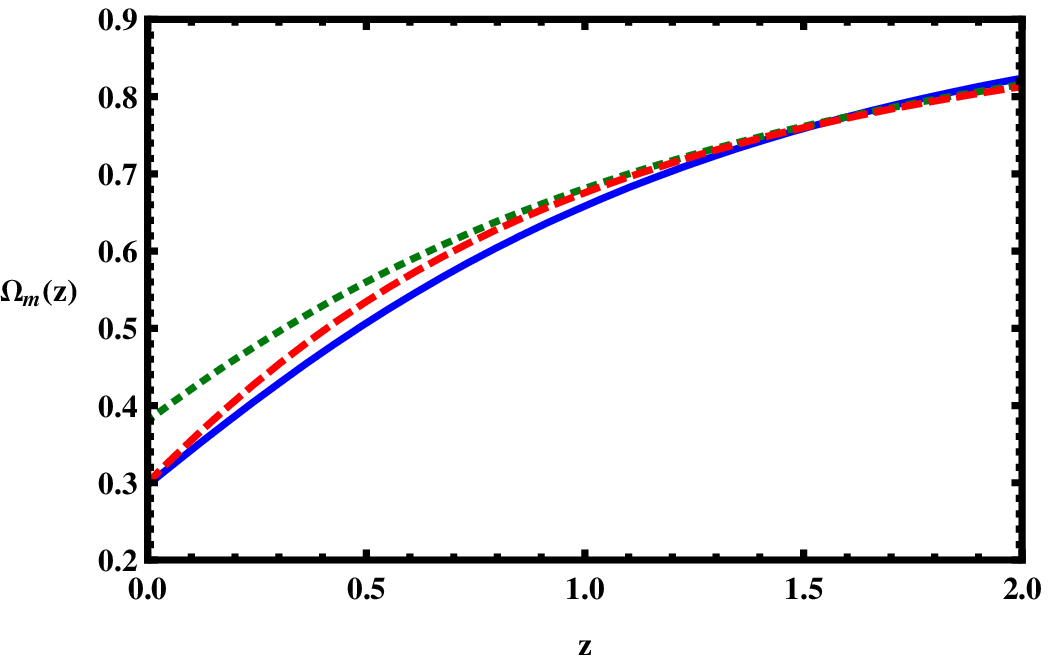, width=6.5cm}
\psfig{file=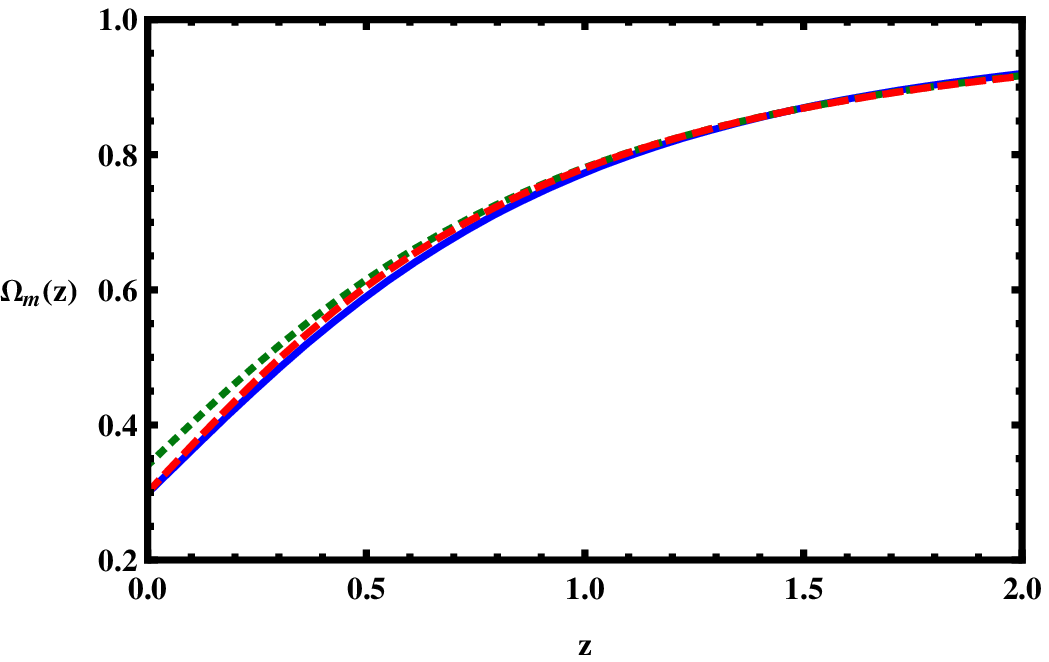, width=6.5cm}
} \vspace{-0.1cm} \caption{
The evolutions of $\Omega_{m}(z)$ obtained from $\omega_{i}^{\ast}$ (dotted), $\oo + \oa \fr{z}{1+z}$ (dashed), and $\omega_{i}$ (solid).
a) When ($\oo, \oa$) = ($-0.9, 0.5$). b) For ($\oo, \oa$) = ($-1.1, 0.3$).} \label{fig1}
\end{figure}
\end{center}

\begin{center}
\begin{figure}
\vspace{1.5cm} \centerline{ \psfig{file=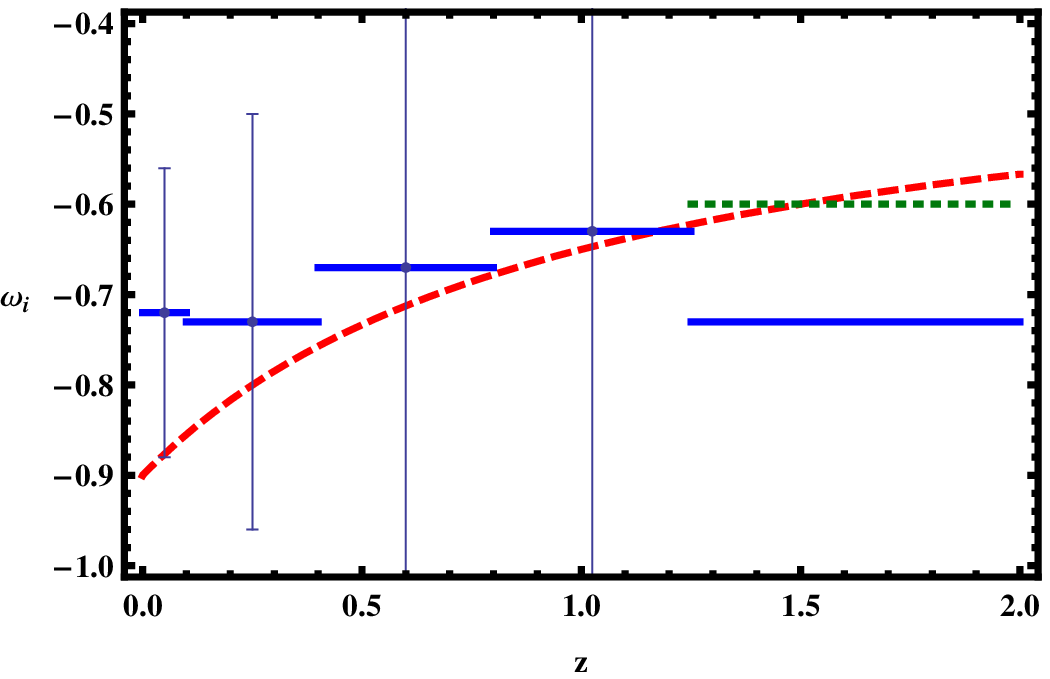, width=6.5cm}
\psfig{file=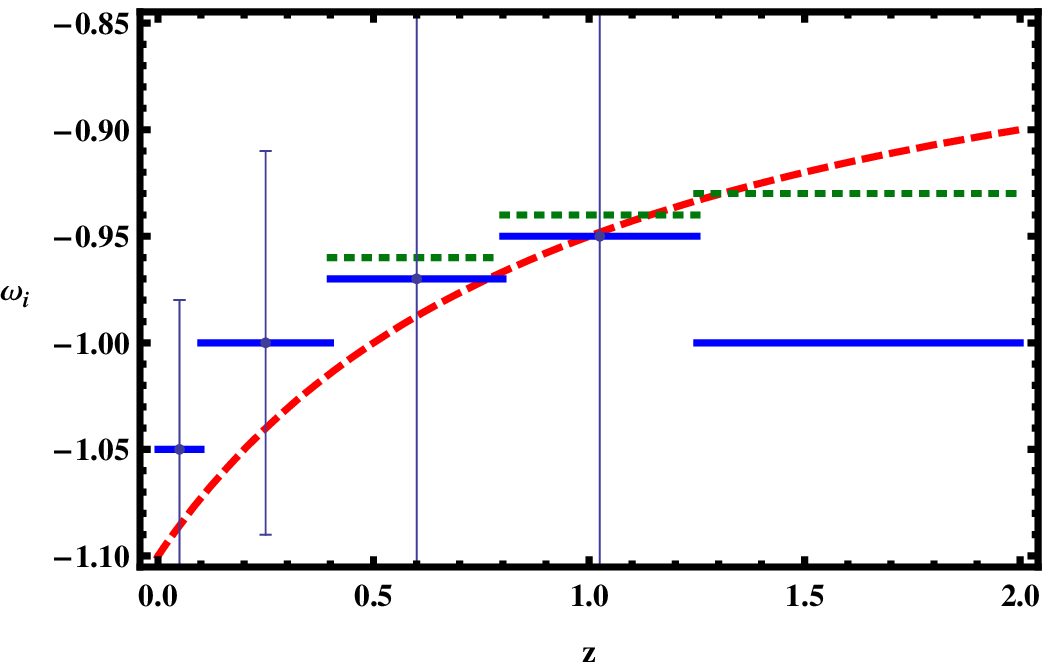, width=6.5cm}
} \vspace{-0.1cm} \caption{
Comparison between $\omega$, $\omega_{i}$, and $\omega_i^{\ast}$.
a) True model is $\omega = -0.9 + 0.5 \fr{z}{1+z}$ (dashed) and the obtained values of $\omega_{i}$ from PCA with the constraint (solid) and without the constraint on the last component (dotted). b) True model is $\omega = -1.1 + 0.3 \fr{z}{1+z}$ (dashed) and the obtained values of $\omega_{i}$ from PCA with the constraint (solid) and without the constraint on the last component (dotted).} \label{fig2}
\end{figure}
\end{center}

Even though, the above constrain Eq. (\ref{omegaj}) seems to be trivial but the piecewise parametrization of $\omega$ suffers the inconsistency without this. For example, one obtains the best fit values of $\omega_i$ from the $\chi^2$ fitting after $z_i$, $z_j$, $\Omega_{m}^{0}$, and $H_{0}$ are specified. Without the above constrain Eq. (\ref{omegaj}), one is not able to recover the original $\Omega_{m}^{0}$ value which is used in the maximum likelihood analysis. This is true whether one uses the marginalization of nuisance parameter $\Omega_{m0}$ in the analysis or not. We show this in Fig. \ref{fig1} by using $\omega_i$ and $z_i$ values in table \ref{table1}. When one use the raw obtained value of $\omega_{i}^{\ast}$ (dotted lines), $\Omega_{m}^{0}$ is different from the one used in the analysis as shown in the Fig. \ref{fig1}. The evolutions of $\Omega_{m}(z)$s of the true models are depicted as the dashed lines. The evolutions of $\Omega_{m}(z)$s obtained from the correct consideration of degree of freedom are shown as the solid lines. The left and right panels correspond to ($\oo$, $\oa$) = ($-0.9, 0.5$) and ($-1.1, 0.3$), respectively.
%%%%%%%%%%%
%%%%%%%%%%%%%%

We find several drawbacks in PCA. Firstly, when the time variation of EOS is not negligible, PCA can not produce the proper behavior of EOS at the entire region of the investigated redshifts. Secondly, PCA produces the EOS which is confused with the constant EOS with the improper present value of it when it changes slowly ({\it i.e.} $\oa \le 0.3$). In Fig. \ref{fig2}, we demonstrate both cases. In the left panel of Fig. \ref{fig2}, we show $\omega_{i}$ (solid) and $\omega_{i}^{\ast}$ (dotted) when the true model (dashed) is ($\oo, \oa$) = ($-0.9, 0.5$). The obtained values of $\omega_{i}$ show the oscillation behavior around $-0.7$ even though the true model decreases monotonically. Thus, PCA method produces the totally different behavior of $\omega$ when $\oa$ is non-negligible. Also in the right panel of Fig. \ref{fig2}, the obtained $\omega_{i}$ is almost same as that of the cosmological constant ($\Lambda$) for the entire region of $z$ even though the true model is ($\oo, \oa$) = ($-1.1, 0.3$). We can compare this result with one in Ref. \cite{09083186}. Even though the result in the mentioned reference seems to be consistent with the cosmological constant, there still can be the viable time varying DE models which can mimic $\Lambda$. This impedes any proper interpretation of the result obtained from PCA method. Thus, PCA also may mislead to the true property of dark energy in the slow changing $\omega$. We check that PCA method can give the reliable result only when $\omega$ is almost constant.
%%%%%%%%%%%
%%%%%%%%%%%%%%

Even though, PCA is the most model independent method for probing DE, we show that the true degree of freedom of the parameters should be one less than the binned number. We also demonstrate that PCA method may mislead to the property of dark energy when the time variation of it is not negligible. Even for the slowly varying $\omega$, PCA result may produce the incorrect information on the true DE. We check that the above conclusion is same for the increasing $\omega$. PCA is adequate only when $\omega$ is a constant. Thus, we may need to check both CPL like model dependent $\omega$ parametrization and PCA method to investigate the DE properly.

%%%%%%%%%%%%%%%%%%%%%%%%%%%%%%%%%%%%%%%%%%%%%%%%%%%%%%%%%%%%%%%%%%%%%%%%
\section*{Acknowledgments}
%%%%%%%%%%%%%%%%%%%%%%%%%%%%%%%%%%%%%%%%%%%%%%%%%%%%%%%%%%%%%%%%%%%%%%%%%
This work was supported in part by the National Science Council, Taiwan, ROC under the
Grant NSC 98-2112-M-001-009-MY3.

\end{document}